\documentclass[aps,prc,twocolumn,superscriptaddress]{revtex4-1}

\usepackage{graphicx}
\usepackage{amsmath}
\usepackage{slashed}
\usepackage{xcolor}
\usepackage{subfigure,dcolumn}
\usepackage{setspace} 
\usepackage[colorlinks,
linkcolor=blue,
anchorcolor=blue,
urlcolor=red,
citecolor=blue]{hyperref}

\begin{document}
		
\title{Isospin-dependent in-medium nucleon-Delta elastic cross section}
	
\author{Manzi Nan}
\affiliation{Institute of Modern Physics, Chinese Academy of Sciences, Lanzhou 730000, China}
\affiliation{School of Nuclear Science and Technology, University of Chinese Academy of Sciences, Beijing 100049, China}
\affiliation{School of Science, Huzhou University, Huzhou 313000, China}
\author{Pengcheng Li}
\email[Corresponding author, ]{lipch@zjhu.edu.cn}
\affiliation{School of Science, Huzhou University, Huzhou 313000, China}
\author{Yongjia Wang}
\affiliation{School of Science, Huzhou University, Huzhou 313000, China}
\author{Qingfeng Li}
\email[Corresponding author, ]{liqf@zjhu.edu.cn}
\affiliation{School of Science, Huzhou University, Huzhou 313000, China}
\affiliation{Institute of Modern Physics, Chinese Academy of Sciences, Lanzhou 730000, China}
\affiliation{School of Nuclear Science and Technology, University of Chinese Academy of Sciences, Beijing 100049, China}
\author{Wei Zuo}
\affiliation{Institute of Modern Physics, Chinese Academy of Sciences, Lanzhou 730000, China}
\affiliation{School of Nuclear Science and Technology, University of Chinese Academy of Sciences, Beijing 100049, China}

\date{\today}
	
\begin{abstract}
In heavy-ion collisions at intermediate energies, the production and propagation of $\Delta$ particles are crucial to understanding the nuclear equation of state and inferring the properties of nuclear matter at high densities. 
Based on the self-consistent relativistic Boltzmann-Uehling-Uhlenbeck (RBUU) transport theory, by introducing the isovector $\rho$ meson exchange into the effective Lagrangian and adopting the density-dependent coupling constants, the detailed expressions for isospin-dependent in-medium $N\Delta\rightarrow N\Delta$ elastic cross sections $\sigma_{N\Delta}^{*}$ have been calculated. The energy and density dependence of the isospin-related $\sigma_{N\Delta}^{*}$ as well as the total contributions of $\sigma$, $\omega$ and $\rho$ meson fields are analyzed. 
It is found that the total $\sigma_{N\Delta}^{*}$ has a sensitive center-of-mass energy dependence at lower energies while exhibiting a slight increase as the center-of-mass energy increases. The isospin effect between different isospin-separated channels weakens as the energy and/or density increases. The isospin effect on the density- and energy-dependent in-medium $N\Delta$ elastic cross sections is dominantly caused by the delicate balance of the isovector $\rho$ meson exchange.
\end{abstract}

	
\maketitle

\section{Introduction}
The information extraction of the high-density nuclear equation of state (EoS) is a fundamental topic in nuclear physics and astrophysics communities. It has significant implications for comprehending the properties of the nuclear matter under extreme conditions and the structure of dense stars \cite{Togashi:2017mjp,Lattimer:2000kb,Most:2018eaw}. 
Several decades of experimental and theoretical studies 
indicate that the properties of nuclear matter at saturation density ($\rho_{0}\approx0.16~fm^{-3}$) have been well investigated, however, the information about EoS and symmetry energy at high density is still ambiguous \cite{Danielewicz:2002pu,Li:2008gp,Drischler:2021kxf,Zhang:2018bwq,Sorensen:2023zkk}.
The heavy-ion collision (HIC) experiments in terrestrial laboratories are the most effective method to investigate dense nuclear EoS. 
Modern facilities, such as the High Intensity heavy ion Accelerator Facility (HIAF), the Facility for Antiproton and Ion Research (FAIR), the Relativistic Heavy Ion Collider (RHIC), the Nuclotron-based Ion Collider fAcility (NICA), and so on, will provide essential experimental support for the nuclear EoS investigation in high-density regions \cite{Zhou:2022pxl,Senger:2021tlg}.

Due to the systems created in HICs being extremely hot, dense and short-lived, the most common approach for obtaining information on high baryon density nuclear EoS is to compare the detected particles and their information with the simulation results from the hadronic transport models \cite{Xu:2019hqg,Bleicher:2022kcu}. Currently, there exist two frequently used categories of hadronic transport models \cite{TMEP:2022xjg,Aichelin:1991xy,Chen:2004si}: the Boltzmann-Uehling-Uhlenbeck (BUU) model and the quantum molecular dynamics (QMD) model. 
There are three main ingredients in these transport models, the initialization, the mean-field potential, and the collision term \cite{Xu:2019hqg}. 
The different approaches to these ingredients in the transport model, such as the density profile of the initial nucleus \cite{Yang:2021gwa}, the density-dependent symmetry energy \cite{Li:2008gp,Tong:2020dku}, and the nuclear medium effect on the two-body collisions \cite{Li:2005jy}, will certainly influence the information of final-state observables, which are used to compare with the experimental data to extract the information of nuclear EoS. To determine the details of implementation or physical assumptions behind the diverging simulation results and extract reliable constraints on the EoS from HICs, various comparisons of different transport models have been performed over the years \cite{TMEP:2016tup,TMEP:2021ljz,TMEP:2017mex,TMEP:2019yci}.

For HICs, on the one hand, with increasing center-of-mass (c.m.) energy (concomitant with increased density), non-nucleon degrees of freedom (such as resonance state, hyperon, and quark-gluon plasma) will be produced. 
On the other hand, the two-body cross sections in the dense nuclear matter created by HICs are expected to differ from those in free space, influenced by the surrounding particles. And, the $\pi$ and $K$ mesons produced by resonance decay and two-body scatterings from high-density regions are commonly employed as sensitive probes to constrain high-density nuclear EoS \cite{Xiao:2013awa,Fuchs:2005zg}. Thus, the careful treatment of the production and dynamic evolution processes of these particles and related resonance states is crucial to accurately extracting information on the high-density nuclear EoS. 

As for the elastic and inelastic cross sections, as well as the medium modifications on nucleon-nucleon ($NN$) and $\Delta$-related collisions, have been investigated for several decades. Several microscopic calculation approaches have been developed, such as Brueckner theory \cite{Bohnet:1989dzk,Han:2022quc}, Dirac-Brueckner (DB) theory \cite{Li:1993ef}, variational approach \cite{Pandharipande:1992zz}, and relativistic BUU (RBUU) microscopic transport theory \cite{Li:2006ez}. In addition, the parameterized forms of the in-medium cross sections can be derived by comparison of theoretical calculations with experimental data \cite{Cai:1998iv,Li:2022wvu,Wang:2020xgk}. In Ref. \cite{Ghosh:2016hln}, the self-energy of $\Delta$ based on the real-time formalism of thermal field theory at finite temperature and density, and the $\pi N $ cross section with the medium modification have been evaluated. And, the original Walecka model suggested that the bulk of the attraction-repulsion competition of interaction required for a realistic description of nuclear matter, and could be technically obtained from isoscalar $\sigma$- and $\omega$- meson exchanges\cite{Walecka:1974qa}.
In our previous work \cite{Li:2000sha}, based on the self-consistent RBUU theory framework, the density- and isospin-dependent $NN$ elastic cross sections $\sigma_{NN\rightarrow NN}^{*}$ have been investigated by considering the interactions described by exchanges of $\sigma$, $\omega$, and isovector $\rho$ meson, in which the selection of exchanged mesons is consistent with that in Refs.\cite{bo2003medium,Yoshida:1998zza,Lalazissis:2005de,Raduta:2021xiz}.
Further, the parameterized in-medium $NN$ elastic cross section was introduced into the ultrarelativistic quantum molecular dynamics (UrQMD) model \cite{Li:2006ez,Li:2011zzp}. Furthermore, the energy-, density-, and isospin-dependent $NN$ inelastic cross sections $\sigma_{NN\rightarrow N\Delta}^{*}$ and $\Delta$-formation cross sections $\sigma_{N\pi\rightarrow\Delta}^{*}$ are calculated with the help of the RBUU transport theory \cite{Li:2016xix,Li:2017pis}. It was found that the medium modifications on the two-body cross sections will influence the production of $\pi^{+}$ and $\pi^{-}$ mesons, as well as their yield ratio \cite{Godbey:2021tbt,Li:2022icu}.

It is known that in HICs at intermediate energies, after the hard-$\Delta$ production process via $NN \rightarrow N\Delta$, there will exist $N_{\rm{coll}}(N\Delta\rightarrow NN)$, $N_{\rm{coll}}(N\Delta\rightarrow N\Delta)$, $N_{\rm{decay}}(\Delta\rightarrow N\pi)$, and $N_{\rm{coll}}(N\pi\rightarrow \Delta)$ for a relatively long time before freeze-out. In central Au+Au collisions at $E_{\rm{beam}}=1A$ GeV, the percentage of $N_{\rm{coll}}(N\Delta\rightarrow N\Delta)$ to all collisions is about 5$\%$ at 25 fm/$c$ \cite{Liu:2020jbg}.
Moreover, the data for $\pi$ yields from Au+Au collisions at $E_{\rm{beam}}=1.23A$ GeV measured by HADES Collaboration are not well described by the transport models \cite{HADES:2020ver}. 
In-medium $\Delta$ production and $\Delta$-related elastic cross sections are important ingredients for the theoretical description of such reactions using transport models. 
By testing with the UrQMD model, the proportion of $\Delta$-related scattering is already considerable,
and the collective flows of pions, which reflect the information of the dynamic process of HICs, are significantly affected by the in-medium $N\Delta\rightarrow N\Delta$ elastic cross section.
Therefore, the $\Delta$-related cross sections and the medium modifications should be further discussed and considered.
In Ref. \cite{Cui:2019dmk}, $\sigma^{*}_{N\Delta\rightarrow NN}$ was investigated within the one-boson-exchange model, and it was found that the $\Delta$ mass dependence of the momentum of the outgoing $\Delta$ and the $M$-matrix would affect the $\sigma^{*}_{N\Delta\rightarrow NN}$, especially around the threshold energy. In Ref. \cite{Mao:1996zz}, within the self-consistent RBUU framework, the in-medium $N\Delta$ elastic and inelastic cross sections with the densities up to $2\rho_{0}$ were calculated by adopting the constant-type coupling constant of the effective Lagrangian. In this work, within the self-consistent RBUU framework, by adopting the density-dependent coupling constant, the isospin-dependent $N\Delta$ elastic cross section will be calculated by introducing the $\rho$ meson exchange in effective Lagrangian. Here, we focus only on the contribution of the isovector $\rho$ meson field, the contributions of the scalar-isovector $\delta$ meson field and the effects of mass splitting on the $N \Delta$ elastic cross section will be shown and discussed in future work.

The structure of the paper is as follows: Section \ref{sec2} introduces the RBUU equation of the $\Delta$ distribution function and the corresponding formulas for the in-medium isospin-dependent $N \Delta \rightarrow N \Delta$ elastic cross sections. Section \ref{sec3} presents the energy and density dependence of the $N \Delta \rightarrow N \Delta$ elastic cross sections. Finally, Section \ref{sec4} gives the conclusion.
	
\section{Theoretical framework}\label{sec2}
 
In this work, we suppose that the interaction between nucleons and $\Delta$ baryons is described through exchanges of $\sigma$, $\omega $ and $\rho$ mesons, with the isospin vector component of the nuclear force is introduced by exchange of $\rho$ meson. 
The effective Lagrangian density reads as 
\begin{equation}
L=L_{F}+L_{I},
\end{equation}
where $L_{F}$ is the free part and $L_{I} $ is the interaction part of effective Lagrangian density.
\begin{equation}
\begin{aligned}
L_{F}=&  \bar{\Psi}_{}\left[i \gamma_{\mu} \partial^{\mu}-m\right] \Psi_{}+ \overline{\Psi}_{\Delta   \nu}\left[i\gamma_{\mu} \partial^{\mu}-m_{\Delta }\right] \Psi_{\Delta }^{\nu}  \\&+\frac{1}{2} \partial_{\mu} \sigma \partial^{\mu}\sigma-\frac{1}{4} F_{\mu \nu} \cdot F^{\mu \nu}-\frac{1}{4} \boldsymbol{L}_{\mu \nu} \cdot \boldsymbol{L}^{\mu \nu}\\&-\frac{1}{2} m_{\sigma}^{2} \sigma^{2}+\frac{1}{2} m_{\omega}^{2} \omega_{\mu} \omega^{\mu}+\frac{1}{2} m_{\rho}^{2} \boldsymbol{\rho}_{\mu} \boldsymbol{\rho}^{\mu} ,
\end{aligned}
\end{equation}
where $F_{\mu \nu} \equiv \partial_{\mu} \omega_{\nu}-\partial_{\nu} \omega_{\mu} , \boldsymbol{L}_{\mu \nu} \equiv \partial_{\mu} \boldsymbol{\rho}_{\nu}-\partial_{\nu} \boldsymbol{\rho}_{\mu} $. $\psi$ is the Dirac spinor of nucleon, and $\psi_{\Delta}$ is the Rarita-Schwinger field of $\Delta$.
\begin{equation}
\begin{aligned}
L_{I}=&g_{N N}^{\sigma} \bar{\Psi} \Psi \sigma-g_{N N}^{\omega} \bar{\Psi} \gamma_{\mu} \Psi \omega^{\mu}-g_{N N}^{\rho} \bar{\Psi} \gamma_{\mu} \boldsymbol{\tau} \cdot \Psi \boldsymbol{\rho}^{\mu}
\\&+g_{\Delta \Delta}^{\sigma} \bar{\Psi} _{\Delta } \Psi_{\Delta } \sigma-g_{\Delta \Delta}^{\omega} \bar{\Psi} _{\Delta }\gamma_{\mu} \Psi_{\Delta } \omega^{\mu}
\\&-g_{\Delta \Delta}^{\rho} \bar{\Psi}_{\Delta } \gamma_{\mu} \boldsymbol{\tau} \cdot \Psi _{\Delta }\boldsymbol{\rho}^{\mu}.
\end{aligned}
\end{equation}
Here, $g_{N N}^{\sigma}$, $g_{N N}^{\omega}$ and $g_{N N}^{\rho}$ are the coupling constants of $\sigma$, $\omega$, $\rho$ mesons with nucleons, respectively. And $g_{\Delta \Delta}^{\sigma}$, $g_{\Delta \Delta}^{\omega} $ and $g_{\Delta \Delta}^{\rho}$ are the coupling constants of $\sigma$, $\omega$, $\rho$ mesons with $\Delta$, respectively. In this work, the parameter sets of the density-dependent coupling constants DD-ME2 are taken from Ref. \cite{Lalazissis:2005de}, the functional form can be written as
\begin{equation}
g_{i}(\rho)=g_{i}(\rho_{sat})f_{i}(\xi), \quad  i=\sigma,\omega,
\end{equation}
and
\begin{equation}
f_{i}(\xi)=a_{i}\dfrac{1+b_{i}(\xi+d_{i})^{2}}{1+c_{i}(\xi+d_{i})^{2}},~~\xi=\frac{\rho}{\rho_{sat}}.
\end{equation}
The $m_{\Delta}$ = 1232 MeV and $m$ = 938 MeV are adopted, and the coupling constant ratios are defined as: $\chi_{\sigma}=\frac{g_{\Delta \Delta}^{\sigma}}{g_{N N}^{\sigma}},~ \chi_{\omega}=\frac{g_{\Delta \Delta}^{\omega}}{g_{N N}^{\omega}},~\chi_{\rho}=\frac{g_{\Delta \Delta}^{\rho}}{g_{N N}^{\rho}}$.
Hence $\chi_{\sigma}=1.0, \chi_{\omega}=0.8, \chi_{\rho}=0.7$ are adopted \cite{Raduta:2021xiz}. Additionally, the functional form of coupling constant by exchange of $\rho$ meson which reads\cite{Lalazissis:2005de}
\begin{equation}
g_{\rho}\left(\rho\right)=g_{\rho}(\rho_{sat}) e^{-a_{\rho} (\xi-1)}.
\end{equation}

Moreover, compared to the Walecka model, the density-dependent coupling constants obtained from DB interactions or the phenomenological approach provide a more precise representation for characterizing finite nuclei and nuclear matter. Additionally, the effective Lagrangian with density-dependence of coupling constants provides a more realistic description of neutron matter, asymmetric nuclear matter, and finite nuclei than the nonlinear meson-exchange model  \cite{Niksic:2002yp,Hofmann:2000vz}.

In this work, the framework of the RBUU approach developed by our group is adopted, which includes an effective Lagrangian with mesons coupling to both nucleons and  $\Delta$(1232) resonances\cite{Mao:1996zz,Li:2017pis,li2000isospin,Li:2016xix}. Here, only the necessary formulas on the isospin-dependent  $N\Delta\rightarrow N\Delta$ cross sections are provided. And the RBUU transport equation of $\Delta$ distribution function reads as
\begin{eqnarray}
&\{p_{\mu}\left[\partial_{x}^{\mu}-\partial_{x}^{\mu} \Sigma_{\Delta}^{\nu}(x) \partial_{\nu}^{p}+\partial_{x}^{\nu} \Sigma_{\Delta}^{\mu}(x) \partial_{\nu}^{p}\right]\nonumber \\
&+m_{\Delta}^{*} \partial_{x}^{\nu} \Sigma_{\Delta}^{S}(x) \partial_{\nu}^{p}\} \frac{f_{\Delta}(\mathbf{x}, \mathbf{p}, \tau)}{E_{\Delta}^{*}(p)}=C^{\Delta}(x, p).
\end{eqnarray}
Here, we only calculate the elastic part, the collision term reads
\begin{equation}
\begin{split}
& C^{\Delta}(x, p) = \frac{1}{4} \int \frac{d^{} \mathbf{p}_{2}}{(2 \pi)^{3}} \int \frac{d^{} \mathbf{p}_{3}}{(2 \pi)^{3}} \int \frac{d^{} \mathbf{p}_{4}}{(2 \pi)^{3}}  \\
&  \times(2 \pi)^{4} \delta^{(4)}(p_{1}+p_{2} -p_{3}-p_{4}) W^{\Delta}(p_{1}, p_{2}, p_{3}, p_{4})[F_{2}-F_{1}].
\label{eq.8}
\end{split}
\end{equation}
The $W^{\Delta}(p_{1}, p_{2}, p_{3}, p_{4})$ is the transition probability, $F_{2}$ and $F_{1}$ are Uehling-Uhlenbeck Pauli-blocking factors of the loss and gain terms.
The relation of cross section and transition probability can be characterized by the formula
\begin{equation}
\begin{aligned}
\int v \frac{d\sigma^{*}}{d\Omega }  d \Omega=
&\int \frac{d^{} \mathbf{p}_{3}}{(2 \pi)^{3}} \int \frac{d^{} \mathbf{p}_{4}}{(2 \pi)^{3}}(2 \pi)^{4} \delta^{4}(p+p_{2}-p_{3}-p_{4}) \\
& \times W^{\Delta}\left(p, p_{2}, p_{3}, p_{4}\right).
\label{eq.9}
\end{aligned}
\end{equation}
Inserting Eq. \ref{eq.9} into Eq. \ref{eq.8}, we obtain
\begin{equation}
C^{\Delta}(x, p)=\frac{1}{4} \int \frac{d \mathbf{p}_{2}}{(2 \pi)^{3}} \sigma^{\Delta}(s, t) \nu_{\Delta}\left[F_{2}-F_{1}\right] d \Omega.
\end{equation}
And $\sigma^{\Delta}(s,t)$ is $N\Delta$ elastic cross section, $v_{\Delta}$ is the Møller velocity.
The transition probability of $N\Delta \rightarrow N\Delta$ cross section can be written as
\begin{equation}
\begin{array}{l}
W^{\Delta}\left(p, p_{2}, p_{3}, p_{4}\right)=G\left(p, p_{2}, p_{3}, p_{4}\right)+p_{3} \leftrightarrow p_{4} \\
\end{array},
\end{equation}
where
\begin{equation}
G=\frac{g_{\Delta \Delta}^{I} g_{\Delta \Delta}^{J} g_{N N}^{I} g_{N N}^{J} T_{e} \Phi_{e}}{16 E_{\Delta}^{*}(p) E^{*}\left(p_{2}\right) E_{\Delta}^{*}\left(p_{3}\right) E^{*}\left(p_{4}\right)},
\end{equation}
\textit{I} and \textit{J} represent the $\sigma$, $\omega$, $\rho$ meson exchanges, the $T_{e}$ is isospin matrix, the $\Phi_{e}$ is spin matrix.
\begin{equation}
\begin{array}{c}
T_{e}=\left\langle T\left|T_{I}\right| T_{4}\right\rangle\left\langle T_{4}\left|T_{J}\right| T\right\rangle\left\langle t_{6}\left|\tau_{J}\right| t_{5}\right\rangle\left\langle t_{5}\left|\tau_{I}\right| t_{6}\right\rangle. \\
\end{array}
\end{equation} 
%
\begin{eqnarray}
\Phi_{e}=
&&\operatorname{tr}\{\gamma_{A}\left(\slashed{p_{3}}+m_{\Delta}^{*}\right) D^{\nu \mu}\left(p_{3}\right)                                  \nonumber\\
&&\gamma_{B} \operatorname{tr}\left[\gamma_{B^{\prime}}\left(\slashed{p_{2}}+m^{*}\right) \gamma_{A^{\prime}}\left(\slashed{p_{4}}+m^{*}\right)\right]               \nonumber\\
&&\left(\slashed{p}+m_{\Delta}^{*}\right) D_{\mu \nu}(p) D_{AA^{\prime}} D_{BB^{\prime}}\}\nonumber\\
&&\frac{1}{\left(p-p_{3}\right)^{2}-m_{I}^{2}}\frac{1}{\left(p-p_{3}\right)^{2}-m_{J}^{2}},
\end{eqnarray}
here,
\begin{equation}
    D_{\mu \nu}(p) = g_{\mu \nu} - \frac{1}{3} \gamma_{\mu} \gamma_{\nu} - \frac{1}{3m_{\Delta}} (\gamma_{\mu} p_{\nu} - \gamma_{\nu} p_{\mu}) - \frac{2}{3m_{\Delta}^2} p_{\mu} p_{\nu}.
\end{equation}
As shown in Table \ref{table_exchange}, $\textit{A}$, $\textit{B}$, $A^{\prime }$ and $B^{\prime }$ are the subscripts of the gamma matrix. And the scalar-scalar meson exchange ($\sigma-\sigma$) share the same gamma matrix, while the vector-vector meson exchanges ($\omega-\omega$, $\rho-\rho$, and $\omega-\rho$) have the same gamma matrix, as well as that for the scalar-vector meson exchanges ($\sigma-\omega$, $\sigma-\rho$). 
\begin{table}[t]
\centering
\caption{Symbols and notations of $\Phi_{e}$.}
\label{table_exchange}
\begin{spacing}{1.5}
\setlength{\tabcolsep}{2mm}
\begin{tabular}{c|cccc|cc}
\hline\hline
  & $\gamma_{A}$ & $\gamma_{B}$ & $\gamma_{A^{\prime}}$ & $\gamma_{B^{\prime}}$ & $D_{AA^{\prime}}$ & $D_{BB^{\prime}}$ \\
\hline
$\sigma - \sigma$ & 1 & 1 & 1 & 1 & 1 & 1 \\
$\omega-\omega, \rho-\rho$,  $\omega-\rho$ & $\gamma_{\alpha}$ & $\gamma_{\beta}$ & $\gamma_{\alpha^{\prime}}$ & $\gamma_{\beta^{\prime}}$ & $-g^{\alpha\alpha^{\prime}}$ & $-g^{\beta\beta^{\prime}}$ \\
$\sigma - \omega, \sigma - \rho$ & $\gamma_{\alpha}$ & 1 & $\gamma_{\alpha^{\prime}}$ & 1 & $-g^{\alpha\alpha^{\prime}}$ & 1 \\
\hline\hline
\end{tabular}
\end{spacing}
\end{table}

\begin{table}[b]
\centering
\caption{The isospin parameters for individual channels of the $N\Delta\rightarrow N \Delta$ cross section.}
\label{table ablc}
\begin{spacing}{1.5} 
\setlength{\tabcolsep}{2.8mm}
\begin{tabular}{l|cccccc}
\hline\hline
& $d_{1}$ & $d_{2}$ & $d_{3}$ & $d_{4}$ & $d_{5}$ & $ d_{6}$ \\ 
\hline
$p\Delta^{++}(n\Delta^{-})$  & 1  & 1 & 1 & 9/4 & 3/2 & 3/2 \\ 
$n\Delta^{++}(p\Delta^{-})$  & 1  & 1 & 1 & 9/4 & -3/2 & -3/2 \\
$p\Delta^{+}(n\Delta^{0})$  & 1  & 1 & 1 & 1/4 & 1/2 & 1/2 \\ 
$n\Delta^{+}(p\Delta^{0})$  & 1  & 1 & 1 & 1/4 & -1/2 & -1/2 \\
\hline \hline
\end{tabular}
\end{spacing}
\end{table}

The isospin matrices of the individual reaction channels are shown in Table \ref{table ablc}. The terms $d_{1}$-$d_{6}$ represent the $\sigma-\sigma$, $\sigma-\omega$, $\omega-\omega$, $\rho-\rho$, $\sigma-\rho$, and $\omega-\rho$ terms, respectively. For the total reaction channel, it is necessary to average the isospin matrices of individual reaction channels that correspond to the same exchange meson. Thus,
\begin{equation}
\frac{d \sigma_{N \Delta \rightarrow N \Delta}^{*}}{d \Omega}=\frac{1}{(2 \pi)^{2} s} \sum_{i=1}^{6} \frac{A_{i}}{32}\left[d_{i} D_{i}(s, t)+(s, t \leftrightarrow u)\right],
\end{equation}
and the total cross section has the form:
\begin{equation}
\sigma_{N \Delta \rightarrow N \Delta}^{*}=\frac{1}{8} \int d \Omega \frac{d \sigma_{N \Delta \rightarrow N \Delta}^{*}}{d \Omega}.
\end{equation}
For the coupling constants:
\begin{equation}
\begin{array}{l}
A_{1}=g_{N N}^{\sigma}{ }^{2} g_{\Delta \Delta}^{\sigma }{ }^{2} , \quad\quad \quad\quad
A_{2}=g_{N N}^{\omega}{ }^{2} g_{\Delta \Delta}^{\omega }{ }^{2} ,\\ A_{3}=g_{N N}^{\sigma} g_{\Delta \Delta}^{\sigma} g_{N N}^{\omega} g_{\Delta \Delta}^{\omega} ,\quad 
A_{4}=g_{N N}^{\rho}{ }^{2} g_{\Delta \Delta}^{\rho}{ }^{2} , \\
A_{5}=g_{N N}^{\sigma} g_{\Delta \Delta}^{\sigma} g_{N N}^{\rho} g_{\Delta \Delta}^{\rho} , \quad 
A_{6}=g_{N N}^{\omega} g_{\Delta \Delta}^{\omega} g_{N N}^{\rho} g_{\Delta \Delta}^{\rho}.
\end{array}
\end{equation}
$D_{i}$ are the contribution from direct terms, 
\begin{equation}
\begin{aligned}
\mathrm{D}_{1}=&\frac{-1}{\left(t-m_{\sigma}^{2}\right)^{2}}\\
&\left[\frac{\left(4\left(4 m^{* 2}-t\right)\left(t-4 m_{\Delta}^{* 2}\right)\left(18 m_{\Delta}^{* 4}-6 t m_{\Delta}^{* 2}+t^{2}\right)\right)}{\left(9 m_{\Delta}^{* 4}\right)}\right],
\end{aligned}
\end{equation}
\begin{equation}
\begin{aligned}
\mathrm{D}_{2}=& \frac{1}{\left(t-m_{\omega}^{2}\right)^{2}}\left(\frac{1}{\left(9 m_{\Delta}^{* 4}\right)}\right) \\
& 8\bigl\{ 4 m_{\Delta}^{* 4}\left[-2 m^{* 2}(9 s+4 t)+9 m^{* 4}+9 s^{2}+13 s t+3 t^{2}\right] \\
& -4 t m_{\Delta}^{* 2}\left[-m^{* 2}(4 s+t)+2 m^{* 4}+2 s^{2}+3 s t+t^{2}\right] \\
& +t^{2}\left(-4 m^{* 2} s+2 m^{* 4}+2 s^{2}+2 s t+t^{2}\right) \\
& +8 m_{\Delta}^{* 6}\left(9 m^{* 2} - 9 s-t \right) + 36 m_{\Delta}^{* 8} \bigr\},
\end{aligned}
\end{equation}
\begin{equation}
\begin{aligned}
\mathrm{D}_{3} = &\frac{2}{\left(t-m_{\sigma}^{2}\right) \left(t-m_{\omega}^{2}\right) 9 m_{\Delta}^{* 3}} \left\{ 16 m^{*} \left( 18 m_{\Delta}^{* 4} - 6 t m_{\Delta}^{* 2} + t^{2} \right) \right. \\ 
& \left. \left( 2 m^{* 2} + 2 m_{\Delta}^{* 2} - 2 s - t \right) \right\},
\end{aligned}
\end{equation}
\begin{flalign}
&D_{4}=D_{2}\left(m_{\omega} \rightarrow m_{\rho}\right),&\\
&D_{5}=D_{3}\left(m_{\omega} \rightarrow m_{\rho}\right),&
\end{flalign}
\begin{equation}
\begin{aligned}
\mathrm{D}_{6}=& \frac{2}{\left(t-m_{\omega}^{2}\right)\left(t-m_{\rho}^{2}\right)} \left(\frac{1}{\left(9 m_{\Delta}^{* 4}\right)}\right) \\
& 8\left\{4 m_{\Delta}^{* 4}\left[-2 m^{* 2}(9 s+4 t)+9 m^{* 4}+9 s^{2}+13 s t+3 t^{2}\right] \right.\\
& \left.-4 t m_{\Delta}^{* 2}\left[-m^{* 2}(4 s+t)+2 m^{* 4}+2 s^{2}+3 s t+t^{2}\right]+t^{2} \right.\\
&\left.\left(-4 m^{* 2} s+2 m^{* 4}+2 s^{2}+2 s t+t^{2}\right)+8 m_{\Delta}^{* 6}\right.\\ &\left. \left(9 m^{* 2}-9 s-t\right)+36 m_{\Delta}^{* 8}\right\}.
\end{aligned}
\end{equation}
Where
\begin{equation}
s=\left(p_{1}+p_{2}\right)^{2}=\left[E_{\Delta}^{*}(p)+E^{*}\left(p_{2}\right)\right]^{2}-\left(\mathbf{p}+\mathbf{p}_{2}\right)^{2},
\end{equation}
\begin{equation}
\begin{aligned}
t=&\left(p_{1}-p_{3}\right)^{2}=m_{\Delta}^{* 2}+m^{* 2}-\frac{1}{2 s}\left[s^{2}-\left(m_{\Delta}^{* 2}-m^{* 2}\right)^{2}\right] \\
& +2\left|\mathbf{p} \| \mathbf{p}_{3}\right| \cos \theta,\\
u=&\left(p_{1}-p_{4}\right)^{2}=2 m_{\Delta}^{* 2}+2 m^{* 2}-s-t,
\end{aligned}
\end{equation}
\begin{equation}
|\mathbf{p}|=\left|\mathbf{p}_{3}\right|=\frac{1}{2 \sqrt{s}} \sqrt{\left(s-m^{* 2}-m_{\Delta}^{* 2}\right)^{2}-4 m^{* 2} m_{\Delta}^{* 2}}.
\end{equation}

A phenomenological form factor was introduced at individual vertex because of the properties of finite size and the short-range correlation\cite{Mao:1996zz}. The typical form of the nucleon-nucleon-meson vertex reads
\begin{equation}
F_{NNM}(t)=\frac{\Lambda^{2}}{\Lambda^{2}-t}.
\end{equation}
In this work, for the nucleon-nucleon-$\Delta$ vertex, 
$\Lambda_{\Delta}=0.4 \Lambda$ is used, and $\Lambda_{\sigma}=1200~\mathrm{MeV}, \Lambda_{\omega}=808~\mathrm{MeV}, \Lambda_{\rho}=800~\mathrm{MeV}$ \cite{Li:2003vd} are applied for individual meson cut-off mass.

\begin{figure}[b]
\centering
\includegraphics[width=0.95\linewidth]{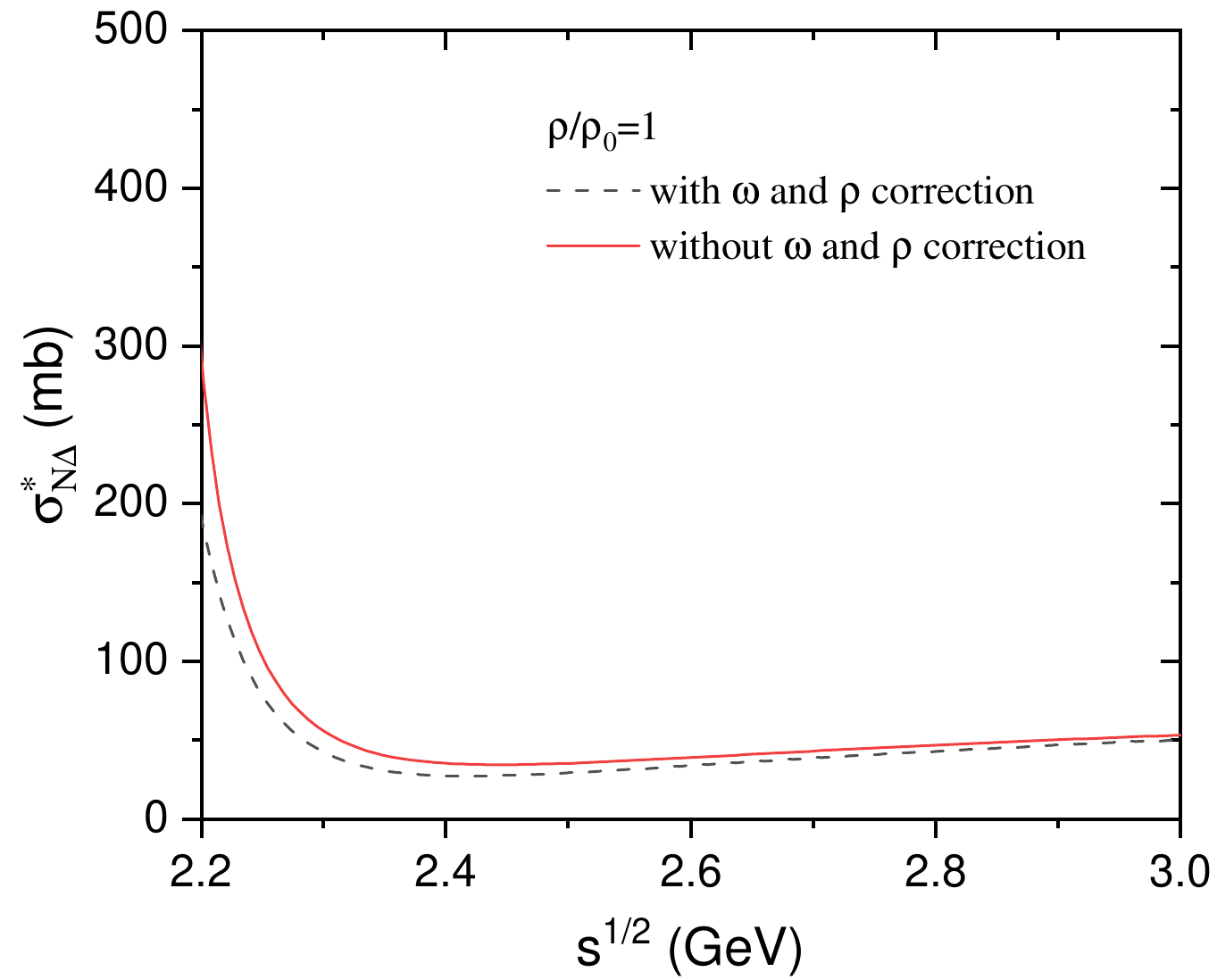}
\caption{The c.m. energy dependence of the in-medium $N\Delta\rightarrow N\Delta$ cross section at normal density, the calculations are performed with and without considering $\omega$ and $\rho$ mesons in the canonical momenta correction.}
\label{fig:graphd}
\end{figure}

In the framework of RBUU theory, the effective mass calculated from the mean-field part serves as an input for the in-medium cross section, which reflects the influence of the potential fields of the surrounding mesons on the properties of single nucleons. With the mean-field approximation, the effective masses of the nucleon and $\Delta$ are only determined by the average value of the $\sigma$ field, 
\begin{eqnarray}
  \mathrm{m}^{*} &=& m + \Sigma_{H}(x), \\
  m_{\Delta}^{*} &=& m_{\Delta}+\Sigma_{H}(x). 
\end{eqnarray}
Consequently, the dependence of the cross section on the mass distribution of a resonance can be achieved by multiplying it by the integral of the Breit-Wigner distribution function\cite{Li:1995pra}. In this work, we focus only on the dependence of the $N\Delta\rightarrow N\Delta$ cross section on the c.m. energy, the baryon density, and especially the isospin. Thus, the transition probabilities of resonance production and absorption are treated as independent of the $\Delta$-mass distribution for simplicity. And, with the help of mass distribution of the $\Delta$ resonance embedded in the microscopic transport model in advance, the calculated in-medium cross section will be further parameterized and introduced into the transport model.

In addition, the in-medium effect on the $N\Delta \rightarrow N \Delta$ cross section should also be associated with the so-called canonical momenta correction \cite{Song:2015hua}. Fig. \ref{fig:graphd} depicts the effects of the canonical momenta correction on the excitation function of the $\sigma^{*}_{N\Delta\rightarrow N\Delta}$ at normal density, the black dotted line and the solid red line represent the calculations with and without considering  $\omega$ and $\rho$ mesons in canonical momenta correction, respectively. It is found that the contribution of the canonical momenta correction on the $\sigma^{*}_{N\Delta\rightarrow N\Delta}$ is relatively weak, especially at c.m. energy above 2.3 GeV. Therefore, it will not be considered in the following calculations.

\section{Results and discussion}\label{sec3}

\begin{figure}\centering
\includegraphics[width=1.0\linewidth]{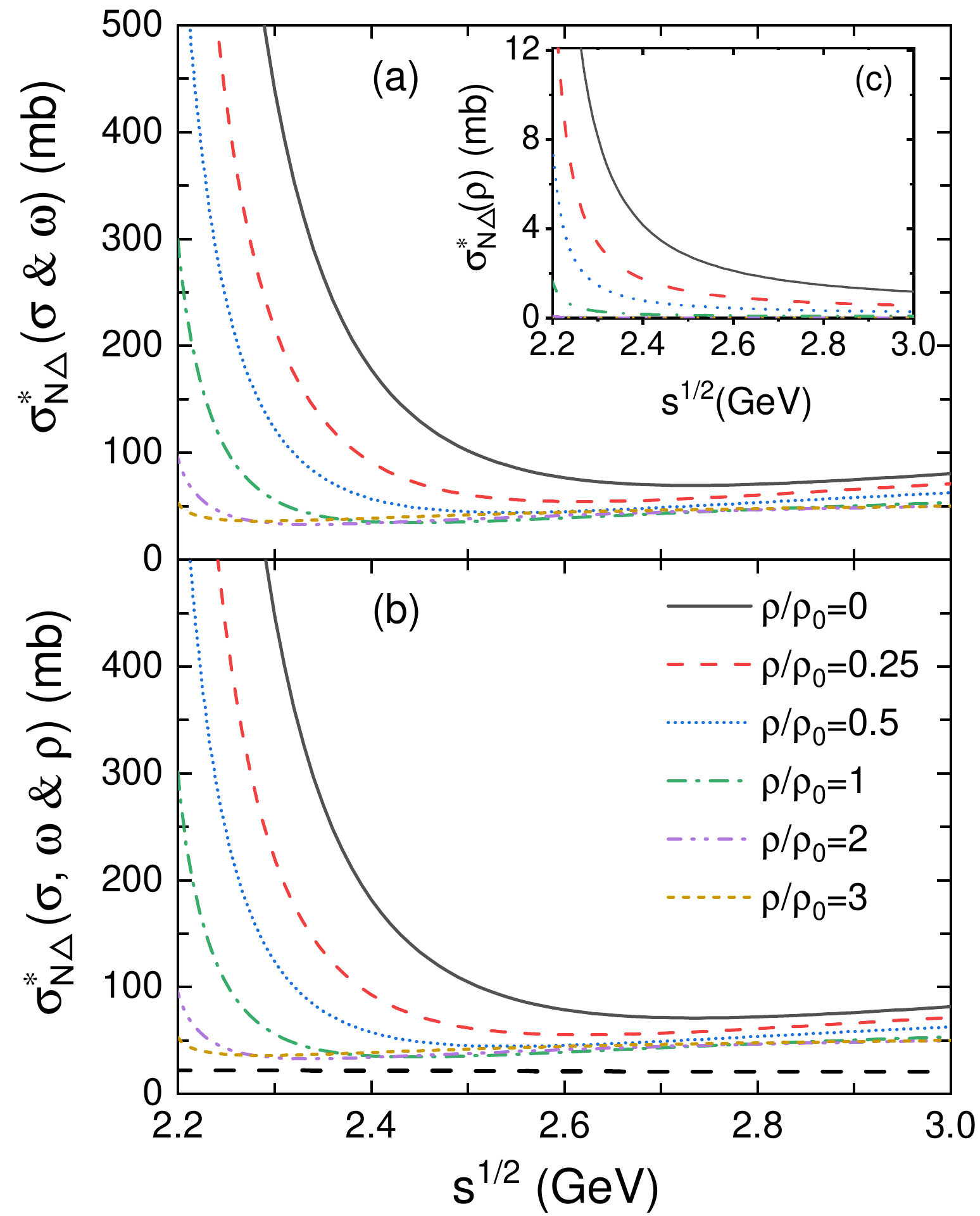}
\caption{(Color online) Panel (a) depicts the contributions of $\sigma$ and $\omega$ meson exchanges to the total cross section $\sigma_{N\Delta}^{*}$ as a function of the center-of-mass energy. Inset panel (c) shows the contribution of $\rho$ meson exchange to $\sigma_{N\Delta}^{*}$. Panel (b) depicts the contributions of $\sigma$, $\omega$ and $\rho$ meson exchanges to total $\sigma_{N\Delta}^{*}$ at various densities.}
\label{fig:graph1}
\end{figure}

\begin{figure*}[htbp]
\centering
\includegraphics[width=1.0\linewidth]{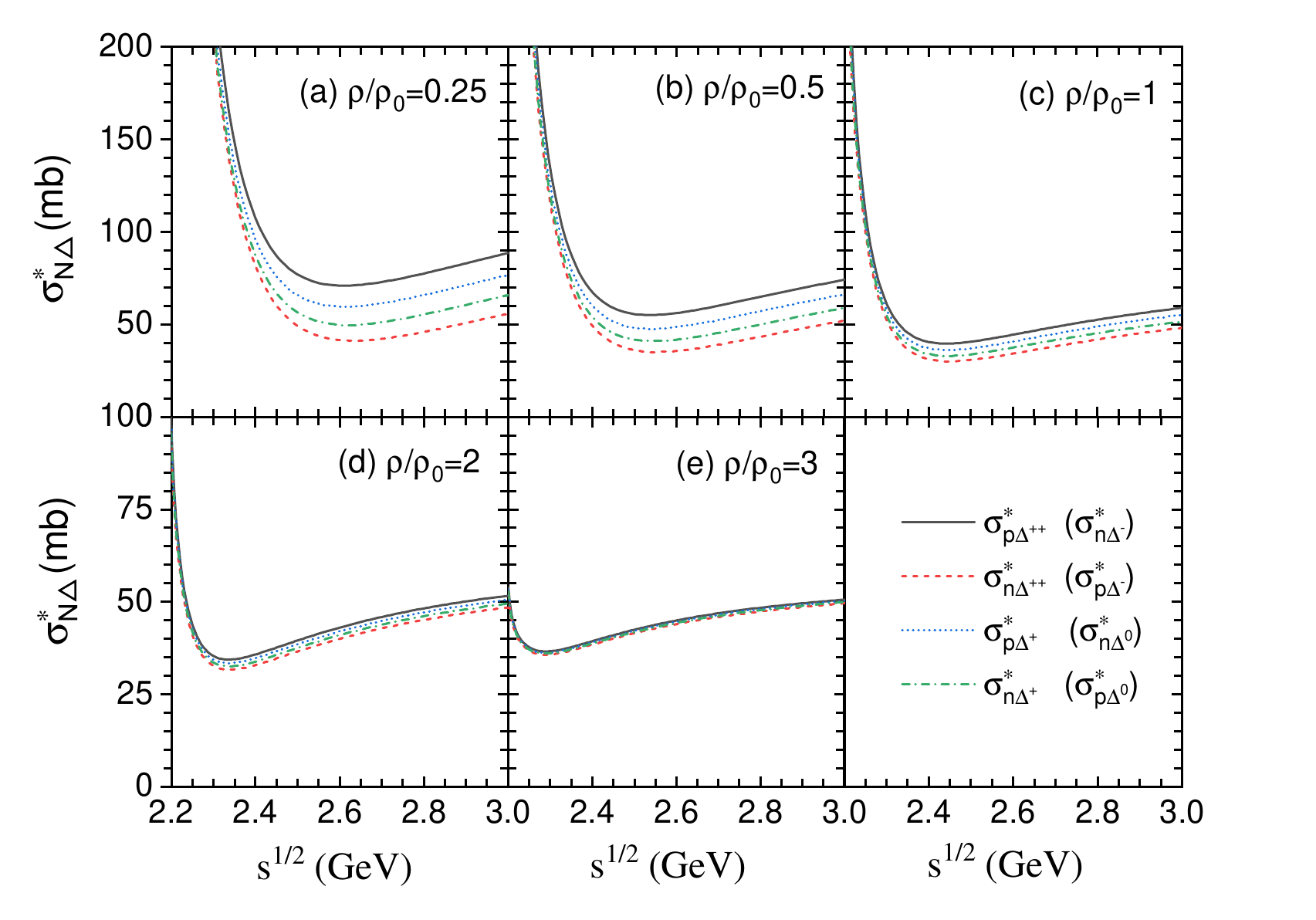}
\caption{The total contribution of all mesons ($\sigma$, $\omega$, $\rho$) for individual in-medium $\sigma^{*}_{N\Delta}$ with densities of 0.25$\rho_{0}$, 0.5$\rho_{0}$, 1$\rho_{0}$, 2$\rho_{0}$, and 3$\rho_{0}$ at the center-of-mass energy of 2.5 GeV.}
\label{fig-2}
\end{figure*}

Firstly, it is of considerable interest to investigate the contributions of $\sigma$, $\omega$, and $\rho$ meson exchanges to the total in-medium $N\Delta\rightarrow N\Delta$ elastic cross sections $\sigma_{N\Delta}^{*}$, as well as their density and center-of-mass energy dependence. 
Fig. \ref{fig:graph1} shows the density-dependent $\sigma_{N\Delta}^{*}$ as a function of the center-of-mass energy at the reduced density varies from 0.0, 0.25, 0.5, 1.0, 2.0 up to 3.0. The top panel shows the contributions of $\sigma$ and $\omega$ meson exchanges (isospin-independent) to $\sigma_{N\Delta}^{*}$, while the inset panel shows the contribution of $\rho$ meson exchange (isospin-dependent) to it. The bottom panel shows the $\sigma_{N\Delta}^{*}$, including the contributions of $\sigma$, $\omega$ and $\rho$ meson exchanges as a function of the center-of-mass energy, the black dashed line is the result of Cugnon’s parametrization. It is found that $\sigma_{N\Delta}^{*}$ decreases with increasing energy and exhibits a slight increase in the high-energy region.  
Furthermore, the isospin-dependent and isospin-independent $\sigma_{N\Delta}^{*}$ decrease with increasing reduced density, and this phenomenon is more pronounced at lower energies. A more detailed analysis will be provided in the following sections. It should be noted that due to the density-dependent coupling constants adopted in this work, the decrease in $\sigma_{N\Delta}^{*}$ with density is faster than that observed in Ref. \cite{Mao:1996zz}, which adopted the constant-type coupling constant.
Furthermore, compared with the total contributions of $\sigma$ and $\omega$ to $\sigma_{N\Delta}^{*}$ (shown in panel (a)), the contribution of $\rho$ meson exchange (shown in panel (c)) is minimal, its contribution approaches zero at 2 to 3 times saturation density.


Subsequently, the total contributions of meson exchanges ($\sigma$, $\omega$, $\rho$) to individual in-medium $N\Delta$ elastic cross sections and their density and center-of-mass energy dependence are explored. 
Fig. \ref{fig-2} shows the center-of-mass energy dependence of the individual $\sigma_{N \Delta \rightarrow N \Delta}^{*}$ at various reduced densities. 
The solid lines represent the $\sigma_{p \Delta^{++}}^{*}(\sigma_{n \Delta^{-}}^{*})$, the short dashed lines represent the $\sigma_{n \Delta^{++}}^{*}(\sigma_{p \Delta^{-}}^{*})$, the short dotted lines represent the $\sigma_{p \Delta^{+}}^{*}(\sigma_{n \Delta^{0}}^{*})$, and the short dashed-dotted lines represent the $\sigma_{n \Delta^{+}}^{*}(\sigma_{p \Delta^{0}}^{*})$. 
It is evident that the splittings in $\sigma_{N \Delta \rightarrow N \Delta}^{*}$ between the individual cross sections of different isospin-separated channels, caused by the isospin effect, decrease with increasing reduced density. 
Furthermore, the individual cross sections of different isospin-separated channels slightly increase with an increase of the center-of-mass energy. This phenomenon is also observed in the $\sigma_{pp(nn)}^{*}$ as a function of the center-of-mass energy \cite{Li:2000sha}. However, it should be noted that the influence of the isospin vector meson exchange on the $N\Delta$ elastic cross section is different from that on the $NN$ elastic scattering, and this is primarily due to the difference in the isospin of the nucleons and $\Delta$ baryons, which consequently influences the calculation of the isospin matrix. 
Furthermore, at low densities ($\rho <2\rho_{0}$), the splitting between the individual cross sections of different isospin-separated channels, such as $\sigma_{p \Delta^{++}}^{*}(\sigma_{n \Delta^{-}}^{*})$ and $\sigma_{n \Delta^{+}}^{*}(\sigma_{p \Delta^{0}}^{*})$, decreases rapidly at lower energies, and then remains relatively constant with increasing center-of-mass energy. 

 
\begin{figure}[htbp]
\centering
\includegraphics[width=1.0\linewidth]{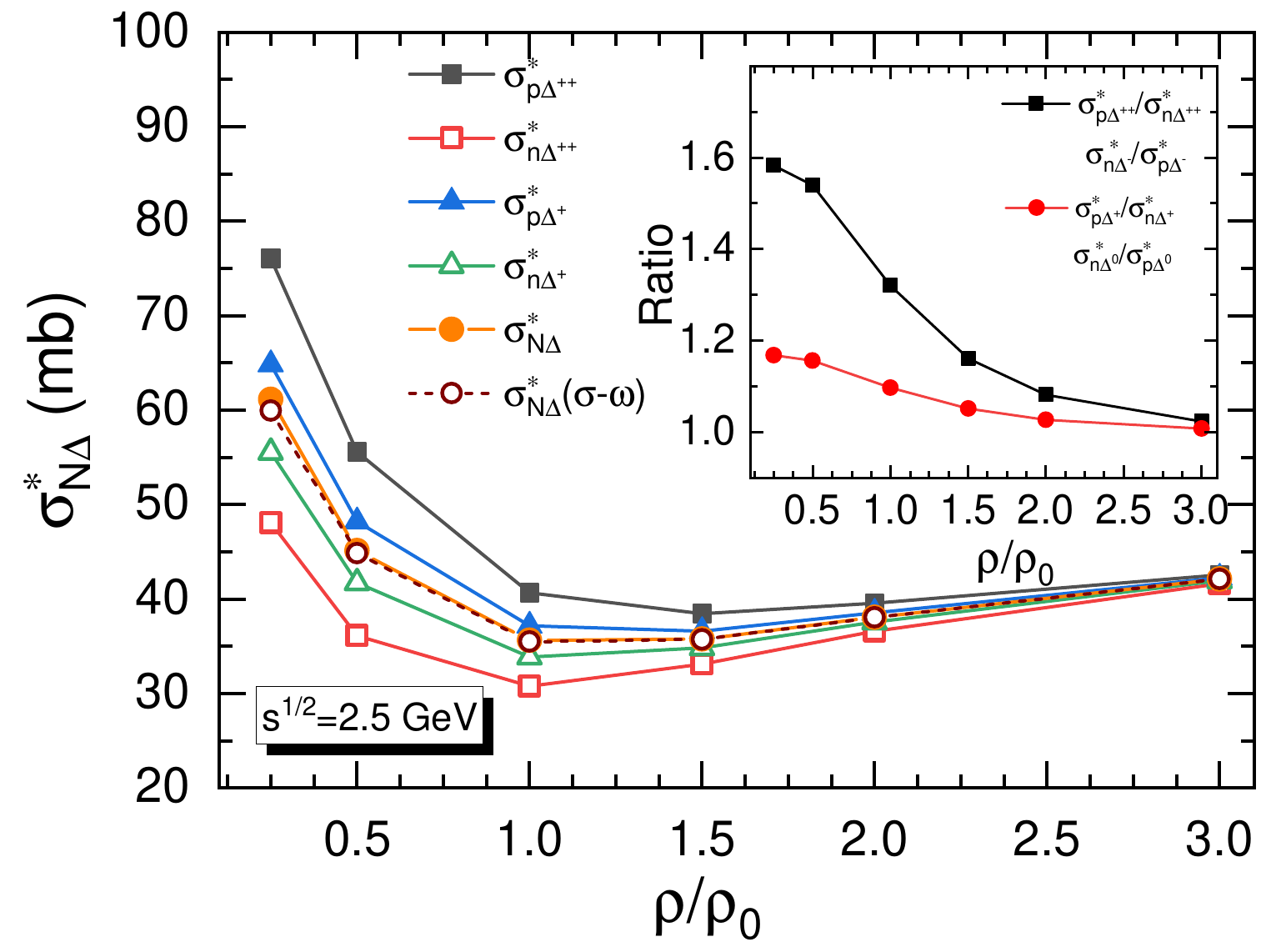}
\caption{(Color online) The reduced density dependence of the individual and total $\sigma_{N\Delta}^{*}$ at a center-of-mass energy of $\sqrt{s}=2.5$ GeV. The inset panel exhibits the ratios of $\sigma_{p\Delta^{++}}^{*}(\sigma_{n\Delta^{-}}^{*})$ to $\sigma_{n\Delta^{++}}^{*}(\sigma_{p\Delta^{-}}^{*})$ and $\sigma_{p\Delta^{+}}^{*}(\sigma_{n\Delta^{0}}^{*})$ to $\sigma_{n\Delta^{+}}^{*}(\sigma_{p\Delta^{0}}^{*})$ as functions of the reduced density.}
\label{fig-3}
\end{figure}

To clearly observe the density dependence of the cross sections of each $N\Delta$ elastic channel, the individual and total $\sigma_{N\Delta}^{*}$ cross sections as functions of the reduced density at $\sqrt{s}=2.5$ GeV are shown in Fig. \ref{fig-3}.
The isospin-independent total $\sigma_{N\Delta}^{*}(\sigma-\omega)$ cross section, which does not include the contribution from $\rho$ meson exchange, is shown by the open circles and serves as the benchmark.
A clear density dependence of the $\sigma_{N \Delta}^{*}$ is observed, and this density dependence gradually weakens with increasing density.
The splitting in the cross section between $p\Delta^{++}$ (solid squares) and $n\Delta^{++}$ (open squares) is more pronounced than that between $p\Delta^{+}$ (solid triangles) and $n\Delta^{+}$ (open triangles). This is because the absolute values of $d_{5}$ and $d_{6}$ of $p\Delta^{++}(n\Delta^{-})$ and $n\Delta^{++}(p\Delta^{-})$ are larger than those of $p\Delta^{+}(n\Delta^{0})$ and $n\Delta^{+}(p\Delta^{0})$. Therefore, $\sigma-\rho$ and $\omega-\rho$ terms have a significant contribution to $p\Delta^{++}(n\Delta^{-})$ and $n\Delta^{++}(p\Delta^{-})$ terms.
The inset panel depicts the ratio of $\sigma_{p\Delta^{++}}^{*}$ to $\sigma_{n\Delta^{++}}^{*}$ (black squares) and the ratio of $\sigma_{p\Delta^{+}}^{*}$ to $\sigma_{n\Delta^{+}}^{*}$ (red circles). These ratios decrease as density increases, indicating that the isospin effect weakens as density increases, and when the density reaches 3$\rho_{0}$, the isospin effect almost disappears.
The difference between $\sigma_{N\Delta}^{*}$ (solid circles) and $\sigma_{N\Delta}^{*}(\sigma-\omega)$ (open circles) reflects the total contribution of $\rho$ meson exchange to the individual cross sections of different isospin-separated channels. Due to the mutual cancellation between the $\sigma-\rho$ and $\omega-\rho$ terms of the isospin matrix, the contribution of the isospin vector is primarily from the $\rho-\rho$ term.
	 
\begin{figure}
\centering
\includegraphics[width=1.0\linewidth]{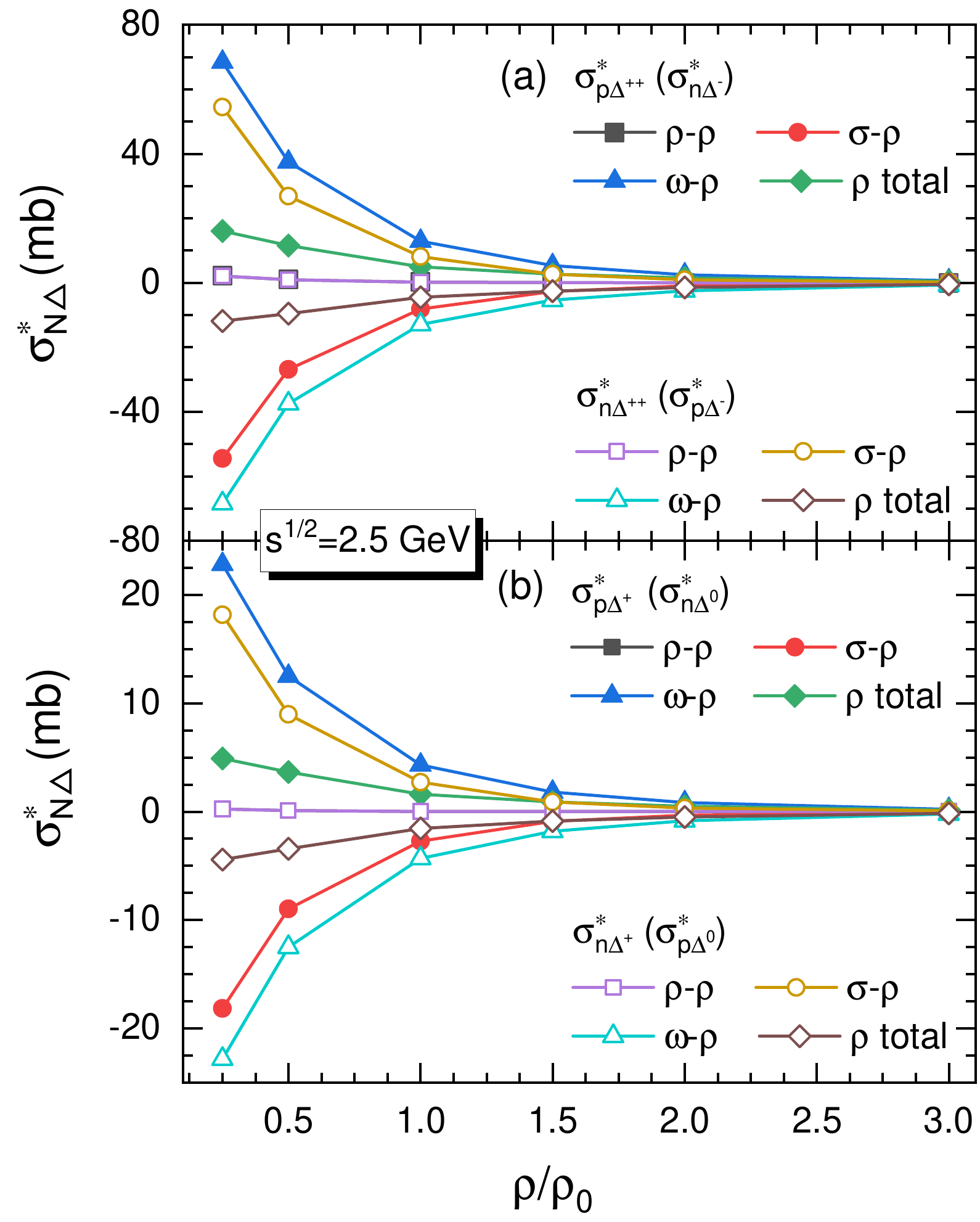}
\caption{The reduced density dependence of $\sigma_{N\Delta}^{*}$, which includes the contribution of $\rho$-related exchange terms, at $\sqrt{s}=2.5$ GeV. The $\sigma_{N\Delta}^{*}$ with three separated exchange terms are shown with squares, circles, and  triangles, respectively, while the $\sigma_{N\Delta}^{*}$ with the total contributions from $\sigma$, $\omega$, and $\rho$ meson exchanges are shown by diamonds.}
\label{fig-4}
\end{figure}

The individual in-medium $N\Delta$ cross section $\sigma_{N\Delta}^{*}$, which includes contributions from the $\rho$-meson related exchange terms, is shown in Fig. \ref{fig-4}. 
To investigate how the $\sigma$, $\omega$, and $\rho$ meson fields affect the $\sigma_{N\Delta}^{*}$, individual contributions from each term are shown, respectively, and the total contribution is indicated by diamonds for comparison. 
As observed, the magnitude of all individual $\sigma_{N\Delta}^{*}$ decreases with density and eventually approaches zero; this density dependence of results from the density-dependent properties of the baryon-baryon-meson coupling constants. 
Further, the exchange terms of the $\omega$ meson to the $\rho$ meson provide the most significant contribution to the $\sigma_{N\Delta}^{*}$, while the exchange terms of the $\sigma$ meson to the $\rho$ meson provide the second most significant contribution, and the exchange terms of the $\rho$ meson to the $\rho$ meson have only a minor contribution. Thus, one can conclude that the contributions of the $\omega$-$\rho$ and $\sigma$-$\rho$ terms are primarily determined by the $\sigma$ and $\omega$ meson fields. 
Moreover, due to the cancellation between the contributions from different exchange terms, the $\sigma_{N\Delta}^{*}$ which includes the total contributions from $\rho$ meson exchanges has a weak density dependence. 
To examine the details from the three individual contributions which are shown in Fig. \ref{fig-4}, one can find that the contributions from isovector-isovector meson ($\rho-\rho$) exchanges are positive (the interaction is attractive) and are always the same sign for both $\sigma^{*}_{p\Delta^{++}}(\sigma^{*}_{n\Delta^{-}})$ and $\sigma^{*}_{n\Delta^{++}}(\sigma^{*}_{p\Delta^{-}})$, and for $\sigma^{*}_{p\Delta^{+}}(\sigma^{*}_{n\Delta^{0}})$ and $\sigma^{*}_{n\Delta^{+}}(\sigma^{*}_{p\Delta^{0}})$. However, for isoscalar-isovector meson ($\sigma-\rho$ and $\omega-\rho$) exchanges, the contributions are always opposite for $\sigma^{*}_{p\Delta^{++}}(\sigma^{*}_{n\Delta^{-}})$ and $\sigma^{*}_{n\Delta^{++}}(\sigma^{*}_{p\Delta^{-}})$, and for $\sigma^{*}_{p\Delta^{+}}(\sigma^{*}_{n\Delta^{0}})$ and $\sigma^{*}_{n\Delta^{+}}(\sigma^{*}_{p\Delta^{0}})$. 

\section{Summary and Outlook}\label{sec4}

Within the RBUU theoretical framework, by adopting the density-dependent coupling constants in the effective Lagrangian, the isospin-dependent in-medium $N \Delta \rightarrow N \Delta$ cross section $\sigma_{N\Delta}^{*}$ is calculated. The density, energy, and isospin dependence of the individual $\sigma_{N\Delta}^{*}$ of different isospin-separated channels are discussed. It has been observed that the behavior of $\sigma_{N\Delta}^{*}$ is consistent with the $NN$ elastic cross section $\sigma_{NN}^{*}$; specifically, the $\sigma_{N\Delta}^{*}$ decreases as the density increases and shows a sensitive dependence on c.m. energy at lower energies, while exhibiting a slight increase as the c.m. energy increases. The splitting in $\sigma_{N\Delta}^{*}$ between different isospin-separated channels weakens as the energy and/or density increase. 
Furthermore, the contributions of the $\sigma$, $\omega$ and $\rho$ meson exchanges to the isospin-related $\sigma_{N\Delta}^{*}$ are analyzed. The contributions from the isoscalar-isovector meson ($\sigma-\rho$ and $\omega-\rho$) exchanges are opposite for $\sigma^{*}_{p\Delta^{++}}(\sigma^{*}_{n\Delta^{-}})$ and $\sigma^{*}_{n\Delta^{++}}(\sigma^{*}_{p\Delta^{-}})$, and for $\sigma^{*}_{p\Delta^{+}}(\sigma^{*}_{n\Delta^{0}})$ and $\sigma^{*}_{n\Delta^{+}}(\sigma^{*}_{p\Delta^{0}})$, while the contributions from the isovector-isovector meson ($\rho-\rho$) exchanges are positive for these cross sections. 
These results suggest that the isospin effect on the density- and energy-dependent in-medium $N\Delta$ elastic cross sections is predominantly caused by the delicate balance of the isovector $\rho$ meson exchange.

In the next step, the $\delta$ and $\pi$ meson fields will be incorporated into the effective Lagrangian to explore the in-medium $N \Delta \rightarrow N \Delta$ elastic cross section in isospin-asymmetric systems. Furthermore, the canonical momenta of $\rho$ and $\omega$ meson fields will also be considered and discussed. Then, the in-medium cross sections will be parameterized and introduced into the microscopic transport model to investigate the production and propagation of particles at several GeV energies. These efforts are helpful to constrain more reliable conclusions on the nuclear EoS at 2 to 4 times saturation density. 

\begin{acknowledgements}
The work is supported in part by the National Natural Science Foundation of China (Nos. 12335008 and 12075085), the National Key Research and Development Program of China under (Grant No. 2020YFE0202002), a project supported by the Scientific Research Fund of the Zhejiang Provincial Education Department (No. Y202353782), the Fundation of National Key Laboratory of Plasma Physics (Grant No. 6142A04230203). The authors are grateful to the C3S2 computing center in Huzhou University for calculation support.
\end{acknowledgements}


\bibliography{ref.bib}
\end{document}